\documentclass[12pt,preprint]{aastex}

\usepackage{graphicx}
\usepackage{amsmath}

\usepackage{epsfig}
\def\msol{\hbox{\kern 0.20em $M_\odot$}}

\newcommand{\lsol}{\hbox{\kern 0.20em $L_\odot$}}
\newcommand{\g}{\hbox{\kern 0.20em g}}
\newcommand{\gmu}{\hbox{\kern 0.20em g$^{-1}$}}
\newcommand{\kg}{\hbox{\kern 0.20em kg}}
\newcommand{\pc}{\hbox{\kern 0.20em pc}}
\newcommand{\mum}{\hbox{\kern 0.20em $\mu$m}}
\newcommand{\mumd}{\hbox{\kern 0.20em $\mu$m$^{-2}$}}
\newcommand{\cm}{\hbox{\kern 0.20em cm}}
\newcommand{\m}{\hbox{\kern 0.20em m}}
\newcommand{\km}{\hbox{\kern 0.20em km}}
\newcommand{\nm}{\hbox{\kern 0.20em nm}}
\newcommand{\s}{\hbox{\kern 0.20em s}}
\newcommand{\h}{\hbox{\kern 0.20em h}}
\newcommand{\smu}{\hbox{\kern 0.20em s$^{-1}$}}
\newcommand{\srmu}{\hbox{\kern 0.20em sr$^{-1}$}}
\newcommand{\smd}{\hbox{\kern 0.20em s$^{-2}$}}
\newcommand{\an}{\hbox{\kern 0.20em an}}
\newcommand{\anmu}{\hbox{\kern 0.20em an$^{-1}$}}
\newcommand{\yr}{\hbox{\kern 0.20em yr}}
\newcommand{\yrmu}{\hbox{\kern 0.20em yr$^{-1}$}}
\newcommand{\Myr}{\hbox{\kern 0.20em Myr}}
\newcommand{\Mymu}{\hbox{\kern 0.20em Myr$^{-1}$}}
\newcommand{\K}{\hbox{\kern 0.20em K}}
\newcommand{\pcmu}{\hbox{\kern 0.20em pc$^{-1}$}}
\newcommand{\pcmd}{\hbox{\kern 0.20em pc$^{-2}$}}
\newcommand{\pcmt}{\hbox{\kern 0.20em pc$^{-3}$}}
\newcommand{\kms}{\hbox{\kern 0.20em km\kern 0.20em s$^{-1}$}}
\newcommand{\kmpd}{\hbox{\kern 0.20em km$^{2}$}}
\newcommand{\kpc}{\hbox{\kern 0.20em kpc}}
\newcommand{\cms}{\hbox{\kern 0.20em cm\kern 0.20em s$^{-1}$}}
\newcommand{\erg}{\hbox{\kern 0.20em erg}}
\newcommand{\ergs}{\hbox{\kern 0.20em erg}}
\newcommand{\cmpd}{\hbox{\kern 0.20em cm$^2$}}
\newcommand{\cmmd}{\hbox{\kern 0.20em cm$^{-2}$}}
\newcommand{\cmms}{\hbox{\kern 0.20em cm$^{-6}$}}
\newcommand{\cmpt}{\hbox{\kern 0.20em cm$^3$}}
\newcommand{\cmmt}{\hbox{\kern 0.20em cm$^{-3}$}}
\newcommand{\mpd}{\hbox{\kern 0.20em m$^2$}}
\newcommand{\mmd}{\hbox{\kern 0.20em m$^{-2}$}}
\newcommand{\mpt}{\hbox{\kern 0.20em m$^3$}}
\newcommand{\mmt}{\hbox{\kern 0.20em m$^{-3}$}}
\newcommand{\mujy}{\hbox{\kern 0.20em $\mu$Jy}}
\newcommand{\mjy}{\hbox{\kern 0.20em mJy}}
\newcommand{\Mj}{\hbox{\kern 0.20em MJy}}
\newcommand{\jy}{\hbox{\kern 0.20em Jy}}
\newcommand{\ghz}{\hbox{\kern 0.20em GHz}}
\newcommand{\G}{\hbox{\kern 0.20em G}}
\newcommand{\muG}{\hbox{\kern 0.20em $\mu$G}}

\newcommand{\thco}{\hbox{${}^{13}$CO}}

\newcommand{\htwo}{\hbox{H${}_2$}}

\newcommand{\water}{\hbox{H$_{2}$O}}

\begin{document}
\title{The CHESS survey of the L1157-B1 shock region~: CO spectral signatures of jet-driven bowshocks}
\author{B. Lefloch$^{1,2}$, S. Cabrit$^{3}$, G. Busquet$^{4}$, C. Codella$^{5,1}$, C. Ceccarelli$^1$, J. Cernicharo$^2$, J.R. Pardo$^2$, M. Benedettini$^{4}$, D.C. Lis$^6$, B. Nisini$^7$}
\altaffiltext{1}{UJF-Grenoble 1 / CNRS-INSU, Institut de Plan\'etologie et d'Astrophysique de Grenoble (IPAG) UMR 5274, Grenoble, F-38041, France\\ lefloch@obs.ujf-grenoble.fr}
\altaffiltext{2}{Centro de Astrobiologia, INTA, Ctra de Torrej\'on a Ajalvir, km 4,
E-28850 Torrej\'on de Ardoz, E-28850 Madrid, Spain}
\altaffiltext{3}{Observatoire de Paris, LERMA, UMR 8112 du CNRS,  ENS, UPMC, UCP, 61 Av. de l'Observatoire, F-75014 Paris, France}
\altaffiltext{4}{ INAF – Istituto di Astrofisica e Planetologia Spaziali, Via Fosso del Cavaliere 100, 00133  Roma, Italy }
\altaffiltext{5}{INAF, Osservatorio Astrofisico di Arcetri, Largo Enrico Fermi 5, I-50125 Firenze, Italy}
\altaffiltext{6}{California Institute of Technology, Cahill Center for Astronomy and Astrophysics 301-17, Pasadena, CA 91125, USA}
\altaffiltext{7}{INAF – Osservatorio Astronomico di Roma, Via di Frascati 33, 00040 Monte Porzio Catone, Italy}

\date{Received~: 2012 July 03; Accepted~: 2012 August 20}
\begin{abstract}
The unprecedented sensitivity of {\em Herschel} coupled with the high resolution of the HIFI spectrometer permits studies of the intensity-velocity relationship  $I$($v$) in molecular outflows, over a higher excitation range than possible up to now. In the course of the CHESS Key Program, we have observed toward the bright bowshock region L1157-B1 the CO rotational transitions between $J$=5--4 and $J$=16--15 with HIFI, and the $J$=1--0, 2--1 and 3--2  with the IRAM-30m and the CSO telescopes. We find that all the line profiles $I_{\rm CO}(v)$ are well fit by a linear combination of three exponential laws $\propto \exp(-|v/v_0|)$ with $v_0= 12.5$, 4.4 and $2.5\kms$. The first component dominates the CO emission at $J \geq 13$, as well as the high-excitation lines of SiO and  $\water$. The second component dominates for $3 \le J_{\rm up} \le 10$ and the third one for
$J_{\rm up} \le 2$. We show that these exponentials are the signature of quasi-isothermal shocked gas components~: the impact of the jet against the L1157-B1 bowshock ($T_{\rm k}\simeq 210\K$), the walls of the outflow cavity associated with B1 ($T_{\rm k}\simeq 64\K$) and the older cavity L1157-B2 ($T_{\rm k}\simeq 23\K$), respectively. Analysis of the CO line flux in the Large-Velocity Gradient approximation further shows that the emission arises from dense gas  ($n(\htwo) \geq 10^5-10^6\cmmt$) close to LTE up to $J$=20.
We find that the CO $J$=2--1 intensity-velocity relation observed in various other molecular outflows is satisfactorily fit by similar exponential laws, which may hold an important clue to their entrainment process.
\end{abstract}

\keywords{ISM: jets and outflows --- ISM: molecules --- Stars: formation}

\section{Introduction}

During the earliest protostellar stages of their evolution, young
stars generate fast collimated winds which impact against the parent cloud through shock fronts,
generating slow "molecular outflows" of swept-up material. The intensity-velocity relationship $I_{\rm CO}$($v$)  observed in low-$J$ CO lines in molecular outflows
has been studied by various authors, as a possible test for discriminating between
entrainment mechanisms. Downes \& Cabrit (2003; hereafter DC03) showed that  hydrodynamical simulations of jet-driven molecular outflows could successfully account for the observed relation $I_{\rm CO}(v)$ in CO $J$=2--1.
The sensitivity and the range of excitation conditions explored were somewhat limited, however.

The heterodyne instrument, HIFI, onboard Herschel\footnote{Herschel is an ESA space
observatory with science instruments provided by European-led principal Investigator
  consortia and with important participation from NASA.}
now allows studies with unprecedented sensitivity, of the
dynamical evolution of gas in protostellar outflows and shocks at spectral and angular resolutions
comparable to the largest ground-based single-dish telescopes (de Graauw et al. 2010). In particular, HIFI  gives access to the CO ladder from $J$=5--4 up to $J$=16--15, probing a wide range of physical conditions.

As part of the CHESS Key Program dedicated to chemical surveys of star forming regions
(Ceccarelli et al. 2010), the outflow shock region L1157-B1 was investigated with Herschel. The protostellar outflow driven by the Class~0 protostar L1157-mm
($d$= 250 pc; Looney et al. 2007) is the prototype of chemically rich bipolar outflows
(see Bachiller et al. 2001 and references therein).  Gueth et al. (1996) showed that the southern lobe of this molecular outflow consists of two cavities, likely created by the propagation of large bowshocks due to episodic events in a precessing, highly collimated jet. Located at the apex of the more recent cavity, the bright bowshock region B1 has been widely studied at millimeter and
far-infrared wavelengths and has become a benchmark for magnetized shock models (see Gusdorf et al. 2008). Preliminary results (Codella et al. 2010) have confirmed the chemical richness of L1157-B1 and revealed the presence of multiple components with different excitation conditions coexisting in the B1 bowshock structure (Lefloch et al. 2010,  Benedettini et al. 2012).

In this Letter, we report on high-sensitivity CO observations with HIFI
of L1157-B1, from $J$=5--4 up to 16--15, and complementary observations
of the $J$=1--0, 2--1 and $J$=3--2 with the IRAM 30m and the CSO telescope.

\section{Observations and data reduction}

\subsection{The HIFI data}
CO transitions between $J$=5--4 and $J$=16--15  were observed with HIFI at the position of L1157-B1 $\alpha_{J2000} = 20^h 39^m 10.2^s$  $\delta_{J2000} = +68^{\circ} 01\arcmin 10.5\arcsec$. The observations were carried out in double beam switching mode. The receiver was tuned in double sideband and the Wide Band Spectrometer (WBS) was used, providing a spectral resolution of 1.1~MHz, which was subsequently degraded to reach a final velocity resolution of $0.5\kms$. The telescope parameters (main-beam efficiency $\eta_{mb}$, half power beamwidth HPBW) were adopted from Roelfsema et al. (2012; see Table~1).

The data were processed with the ESA-supported package
HIPE~6\footnote{HIPE is a joint development by the Herschel Science
  Ground Segment Consortium, consisting of ESA, the NASA Herschel
  Science Center, and the HIFI, PACS and SPIRE consortia.}  (Herschel
Interactive Processing Environment). FITS files from level 2 data were then
created and transformed into
GILDAS\footnote{http://www.iram.fr/IRAMFR/GILDAS} format for baseline
subtraction and subsequent data analysis.

\subsection{Complementary ground-based observations}
The CO $J$=3--2 line emission was mapped at the Nyquist {\bf spatial} frequency across a region of $72\arcsec \times 120\arcsec$  in the southern lobe of the L1157 outflow  in June 2009 using
the facility receivers and spectrometers of the Caltech Submillimeter Observatory (CSO) on Mauna Kea, Hawaii. Observations were carried out in position switching mode using a reference position 10$^{\prime}$ East from the nominal position of B1. Small contamination from the cloud was observed, resulting in a narrow dip at the cloud velocity $v_{lsr}= +2.6\kms$ (Bachiller \& Perez-Gutierrez, 1997). The data were taken under good to average weather conditions, with system temperatures $\rm T_{sys}$ in the range $750-850\K$. An FFTS was used as a spectrometer, which provided a nominal resolution of $0.1\kms$.
The final resolution was degraded to $0.25\kms$.  The final rms of the map is $\approx 0.25\K$ per $0.25\kms$ velocity interval.

Deep integrations  were performed at the frequency of the CO $J$=1--0 and $J$=2--1 transitions
in June and August 2011 as part of an unbiased spectral survey of L1157-B1 at the IRAM 30m telescope (Lefloch et al. 2012, in prep). The EMIR receivers were connected to the 200~kHz resolution (FTS) spectrometers. Observations were carried out using a nutating secondary with a throw of $3\arcmin$, resulting in a narrow absorption feature at the cloud velocity.

The observations (frequency, $\rm Obs\_Id$) and the telescope parameters are summarized in Table~1.  Line intensities are expressed in units of antenna temperature  corrected for atmospheric attenuation (for ground-based observations)
$T_A{}^{*}$.

\section{Results}
\begin{figure}
\includegraphics[width=\columnwidth]{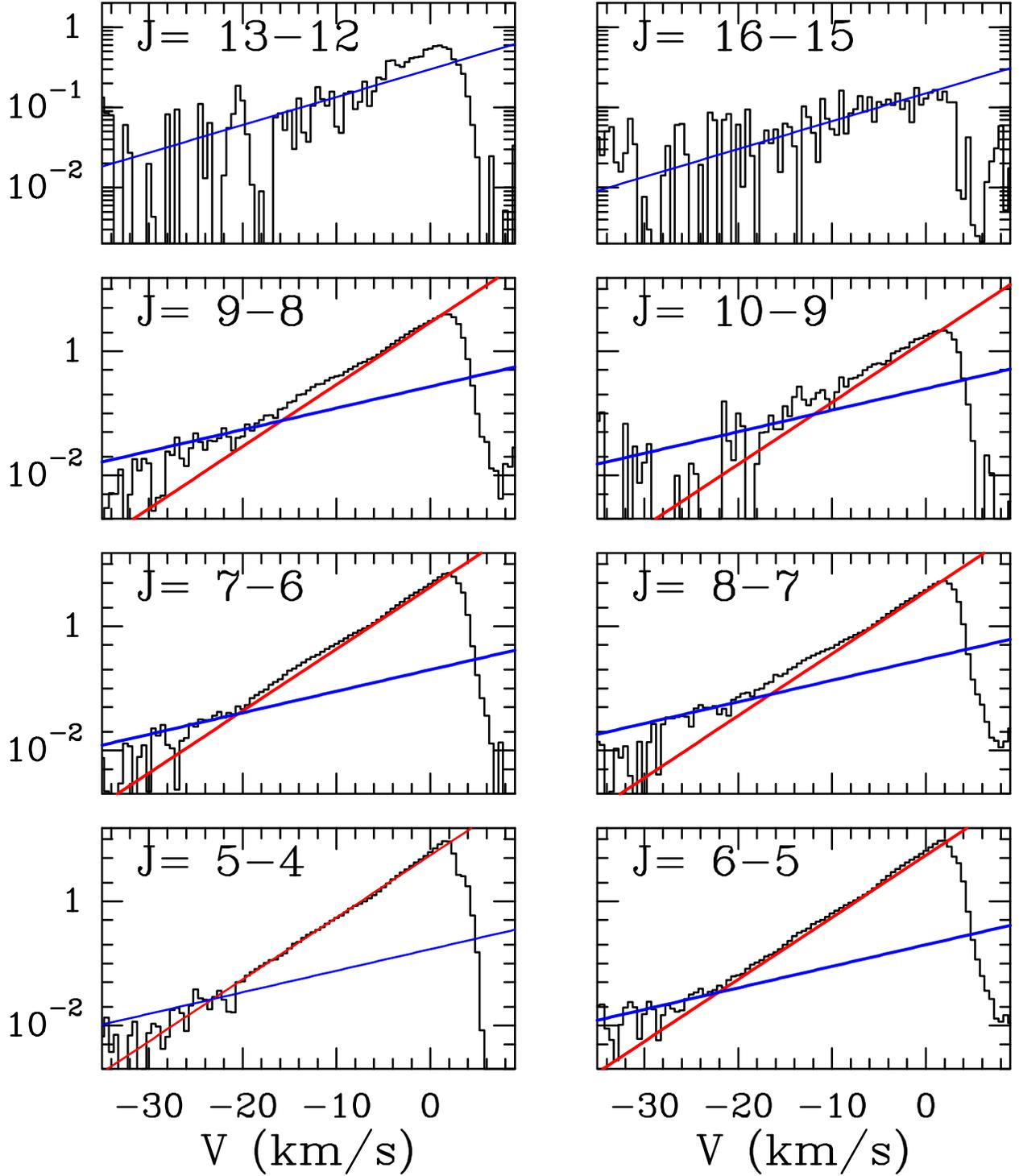}
\caption{Montage of CO line profiles detected with HIFI. The line intensity  is expressed on a logarithmic scale in corrected antenna temperature ($T_A{}^{*}$). The CO profiles are fit by a linear combination of two exponential functions $g_1\propto \exp(-|v/12.5|)$ (blue) and $g_2\propto exp(-|(v/4.4|)$ (red).
}
\end{figure}

\begin{figure}
\includegraphics[width=1.3\columnwidth]{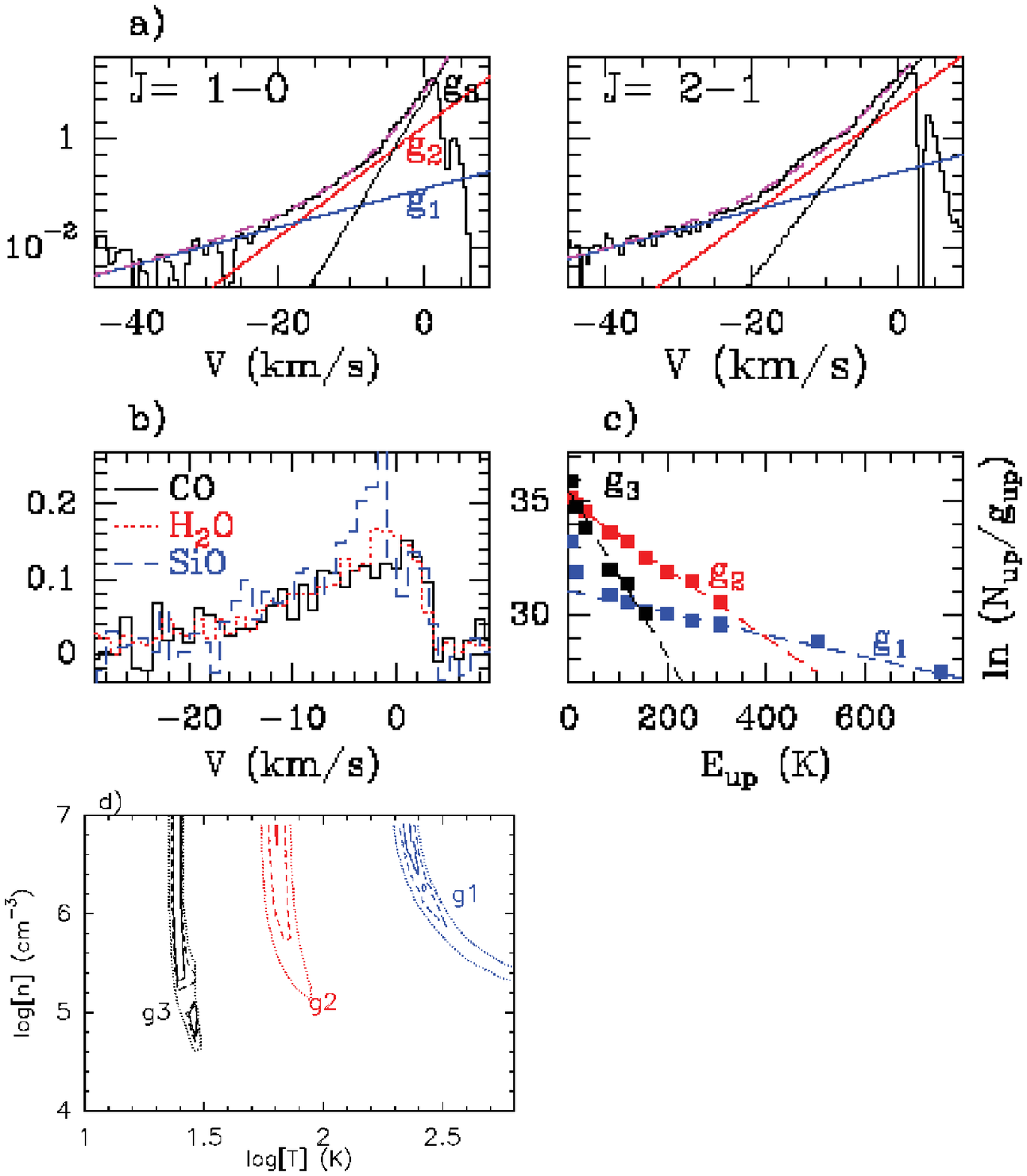}
\caption{a)~CO $J$=1--0 and $J$=2--1 line profiles observed toward L1157-B1 at IRAM. The line intensity  is expressed on a logarithmic scale in corrected antenna temperature ($T_A{}^{*}$). The CO profiles are fit by a linear combination of three exponential functions $g_1\propto exp(-|v/12.5|)$ (blue) and
$g_2\propto \exp(-|v/4.4|)$ (red) and $g_3\propto \exp(-|v/2.0)|$ (black). We superpose the fit to the full CO $J$=1--0 and $J$=2--1 emission with a dashed, magenta line. b)~Similarity of the CO $J$=16--15 ($E_{up}=750\K$; thick black), $\water$ $3_{12}-3_{03}$ ($E_{up}= 215.2\K$; red dotted) and SiO $J$=8--7 line profiles ($E_{up}= 75.0\K$; blue dashed). c)~CO rotational diagrams at $20\arcsec$ spatial resolution for the three components~: $g_1$ (blue), $g_2$ (red), $g_3$ (black). d)~$\chi^2$ distribution of LVG slab models for $g_1$, $g_2$ and $g_3$. Contours levels of 0.5, 1.0, and 2.0 are drawn with solid, dashed and dotted lines, respectively.
}
\end{figure}

\subsection{CO Spectral signatures}
The CO line profiles are  displayed in Fig.~1 ($J_{\rm up} \ge 5$), Fig.~2a ($J=$1--0, 2--1) and Fig.~3 ($J$=3--2) on a log-linear scale. This permits identification of  three underlying components in the line profiles, denoted hereafter $g_1$, $g_2$ and $g_3$. Each component is well described by an exponential law $I_{\rm CO}(v)= I_{\rm CO}(0)\exp(-|v/v_0|)$ showing the same slope at all $J$, but differing relative intensities.

The high-excitation CO transitions $J$=13--12 and $J$=16--15 are well fit by the component $g_1$ {\em alone}, with $v_0= 12.5\kms$  (Fig.~1, top). In the lower $J$ transitions, $g_1$ still dominates the emission at high velocity $v \le -20 \kms$ (Figs.~1--2) and a simple scaling to the $J$=16--15 line profile in this velocity range allows us to determine the total $g_1$ contribution to the integrated intensity of each CO line (Table~1).
After removing the contribution of $g_1$, the $J$=10--9 and $J$=9--8 line profiles appear to be well reproduced by the $g_2$ component alone, with $v_0= 4.4\kms$ (Fig.~1). Hence, the same procedure as above was applied to estimate the total contribution of $g_2$ across the CO ladder (Table~1). An emission excess with respect to the $g_2$ and $g_1$ contributions is observed at velocities close to the cloud velocity at
$J \le 7$ and actually dominates the emission in the low excitation transitions $J \leq 2$ (Table~1).  This "residual emission" is well fit by the third exponential function $g_3 \propto \exp(-|v/2.5|)$, except for the transitions $\rm J\leq 2$ for which $v_0$ has a slightly steeper value $\approx 2.0$.

This profile decomposition is justified by the fact the the $^{12}$CO line emission is optically thin, as shown by comparison with \thco\ spectra, except at velocities very close to that of the ambient cloud for the low-$J$ transitions.

The fact that the slopes of $g_1$, $g_2$, and $g_3$  are {\em independent} of the CO transition considered is  quite remarkable, and somewhat unexpected as one would naively assume the temperature gradients in shocked, accelerated gas to alter the shape of $I_{\rm CO}$($v$) depending on the CO rotational level. Instead, it seems that the upper energy of the level only changes the relative importance of each exponential component in the resulting profile. In the following section, we show that this behavior is due to each component probing a distinct spatial region  with almost uniform excitation conditions.

\subsection{Shock origin and physical conditions}

Figures 2a and 1 show that first $g_3$, then $g_2$ and finally $g_1$  dominate at progressively higher $J_{up}$, which implies that the three components trace gas with progressively higher excitation conditions. We note that all three components emit over a wide range of velocities and all  display an emission peak at velocities close to the cloud velocity. Therefore,
their excitation conditions cannot be determined from a simple analysis (e.g. line ratios) in different velocity intervals. Instead, we use our CO profile decomposition, which yields the total flux of each component as a function of $J_{\rm up}$ (Table~1).
The excitation conditions in each component are first obtained from a simple rotational diagram analysis of the HIFI and IRAM CO fluxes, after convolving to a common angular resolution of $20\arcsec$ (Fig.~2c). Further constraints on the kinetic temperature and density of the CO gas are then obtained using a radiative transfer code in the Large Velocity Gradient (LVG) approximation assuming a plane parallel geometry. We used
the \htwo\ collisional rate coefficients of Yang et al. (2010) and built a grid of models with density between $10^4 \cmmt$ and $10^7 \cmmt$ and temperature between $10\K$ and $1000\K$ ($250\K$) to determine the region of minimum $\chi^2$ as a function of density and temperature for $g_1$ ($g_2$ and $g_3$). We adopted a typical line width $\Delta v$ of $10\kms$ ($g_1$), and $5\kms$ for ($g_2$ and $g_3$).

\subsubsection{The $ g_1$ component}

PACS observations of L1157-B1 have shown that the CO $J$=16--15 emission arises from a small ($\approx 7-10\arcsec$) region, which peaks at $\approx 5\arcsec$ North with respect to the nominal position of B1, and  is associated with a partly-dissociative J-type shock in the region where the protostellar jet impacts the cavity (Benedettini et al. 2012). We conclude that $g_1$ is the spectral signature of this shock and refer to this gas component as the " $g_1$ shock" in the subsequent discussion.

As shown in Fig.~2b, the $g_1$ component alone dominates the profiles of the
$\water$ $3_{12}$--$3_{03}$ ($E_{up}=215\K$) and SiO $J$=8--7 ($E_{up}=75\K$) lines, observed respectively with HIFI (Busquet et al., in prep) and the IRAM 30m telescope (Codella et al. in prep). In spite of large differences in their upper level energies, an excellent match is observed at all velocities between those tracers and the CO $J$=16--15 ($E_{up}=751.8\K$),  except in the low-velocity range  of  the SiO $J$=8--7 transition, where a slight excess is observed.
These three transitions are therefore probing the same physical region and the similarity of their line profiles justifies our use of the CO $J$=16--15  as a template for determining the contribution of  $g_1$ to each CO transition.

Given the wide range of CO transitions and frequencies considered, the variations of the coupling of the telescope beam with the (off-centered) compact $g_1$ shock must be taken into account. This was done by convolving a fully sampled map of the SiO J=8-7 emission obtained at the IRAM 30m telescope  (Codella et al., in prep) to the resolution of the HIFI beams of the CO transitions $J$=5--4 up to $J$=8--7. The $J$=16--15 line flux in a $20\arcsec$ beam  was directly obtained by  convolving the PACS image of Benedettini et al. (2012) and the $J$=13--12 line flux was estimated under the assumption that the ratio of $J$=13--12/$J$=16--15  is preserved when degrading the resolution from $14\arcsec$ ($10.7\arcsec$) to  $20\arcsec$. This is supported by the fact that both lines show similar profiles.
Since the IRAM $J$=2--1 and HIFI $J$=16--15 observations have very similar angular resolution (Table~1), the $J$=2--1 line flux was estimated under the same assumption as for $J$=13--12.

The rotational diagram of $g_1$ is shown in Fig.~2c. The level populations in the HIFI range are
well fit by a single rotational temperature $T_{rot}= 206\K$ and a beam-averaged column density  $N({\rm CO})= 2.3\times 10^{15} \cmmd$. The populations of the levels $J_{\rm up}$=1 and
$J_{\rm up}$=2 lie a factor 3--7 above this trend, and could be the signature of a lower excitation component.
LVG calculations were then carried out taking into account the CO line fluxes from $J$=5--4 to $J$=20--19, using the HIFI and PACS data (see Benedettini et al. 2012). The results are shown in Fig.~2d in the form of $\chi^2$ contours. The best-fit solution ($\chi^2= 0.35$) is obtained for
$T_{\rm k}= 210\K$ and $\rm n(\htwo)\ge $ a few $10^6\cmmt$, and a source-averaged column density $N({\rm CO})= 0.9\times 10^{16}\cmmd$ for a source size of $\approx 10\arcsec$. While our results are consistent with  our previous PACS analysis (Benedettini et al. 2012), they favor dense solutions close to LTE with $n(\htwo) \geq 10^6\cmmt$  and kinetic temperatures in the  range $200-300\K$ (Fig.~2d).

\subsubsection{The $g_2$ component}

Our CSO map shows that the CO $J$=3--2 outflow emission is dominated by the $g_2$ component all over the B1 cavity,from the driving protostar L1157-mm down to the bowshock L1157-B1, at the cavity apex (Fig.~3). The exponent of $g_2$ remains constant at all the positions observed. This suggests that the $g_2$ component arises from the shocked gas in the walls of the  B1 cavity.

The excitation conditions in the $g_2$  component  were derived by scaling the CO fluxes at each $J$ to a common angular resolution of $20\arcsec$, following the same procedure as described above. The coupling between the telescope beam and the source was estimated from the convolution of our CO $J$=3--2 map to the resolution of the different HIFI beams.  The data are well fit by
a single rotational temperature $T_{rot}\simeq 64\K$ and a beam-averaged column density  $N({\rm CO})= 4.0\times 10^{16}\cmmd$. Our LVG calculations again favor an  LTE solution with density
above $10^5\cmmt$ and kinetic temperature in the range $60-80\K$ (see Fig.~2d). The best-fit solution
($\chi^2= 0.27$) is obtained for $n(\htwo)\simeq 1.0\times 10^7\cmmt$,
$T_{\rm k}= 64\K$, and a source-averaged column density $N({\rm CO})= 0.9\times 10^{17}\cmmd$ for a typical source size of $20\arcsec$. Such a value of $T_{\rm k}$ is in good agreement with the value derived by Tafalla \& Bachiller (1995) from multi-transition NH$_3$  observations using the VLA.
The bulk of $\rm NH_3$ emission arises from the cavity walls and peaks at the apex of B1 (Tafalla \& Bachiller, 1995), supporting our interpretation of the $g_2$ component as tracing this cavity.

\begin{figure}
\includegraphics[width=\columnwidth]{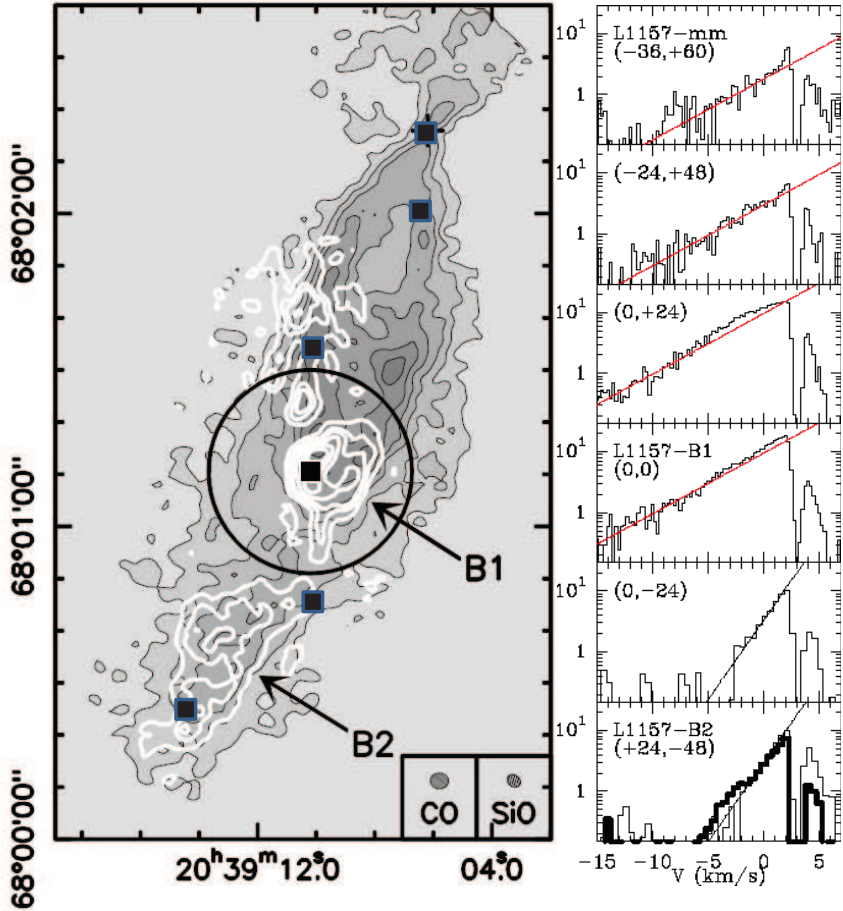}
\caption[]{ (Left)~Southern outflow lobe of L1157 in CO $J$=1--0 (greyscale and
black contours) and in SiO $J$=2--1 (white contours) as observed at the PdBI (Gueth
et al. 1996,1998). The HIFI band 1 FWHM (CO $J$=5--4) is shown as the black circle. Black squares
mark the positions of the CO $J$=3--2 spectra plotted at right.
(Right)~Montage of CO $J$=3--2 spectra at various positions along the outflow between the protostar L1157-mm and the bowshock L1157-B1.  The $g_2$ fit is drawn in red. The black solid line in the spectra of B2 and at offset position (0,$-$24) is representative of $g_3$. The $g_3$ component toward B1 (thick) is superposed on the spectrum of B2  to highlight their similarity.}
\end{figure}

\subsubsection{The $g_3$ component}

The $g_3$ component was identified as the residual emission in the CO line profile, in addition to the contributions of $g_1$ and $g_2$. The CSO data bring some insight into the spatial origin of this component. The CO $J$=3--2 emission from the older B2 cavity, south of B1, is seen to follow the same intensity-velocity distribution $I_{\rm CO}$($v$) $\propto \exp(-|v/2.0|)$ as the $g_3$ component observed toward B1 (see bottom panel in Fig.~3). We thus speculate that $g_3$ is actually tracing shocked gas from the previous ejection, which led to the formation of the B2 outflow cavity.

The excitation conditions in the $g_3$  component  were derived by scaling the CO fluxes at each $J$ to a common angular resolution of $20\arcsec$, following the same procedure as described above.
A rotational diagram analysis of the $g_3$ emission yields $T_{rot}= 26\K$ and a beam-averaged gas column density $N({\rm CO})= 2.6\times 10^{16}\cmmd$.
Our LVG calculations again favor an LTE solution with density $\ge 10^5\cmmt$ and $T_{\rm k}\simeq 23\K$ (Fig.~2d). The best-fit solution was obtained for a source size of $25\arcsec$ and a source-averaged column density $N({\rm CO})\simeq  1.0\times 10^{17}\cmmd$.
The lower temperature of $g_3$ compared to $g_2$ is consistent with the B2 cavity being older than B1, thus having experienced more post-shock cooling
(Gueth et al. (1996) estimated an age of $3000\yr$ and $2000\yr$ for the outflow cavities associated with bowshocks B2 and B1, respectively).

\subsection{The relation $I_{\rm CO}$($v$) revisited}
\begin{figure}[h]
\includegraphics[width=7.cm]{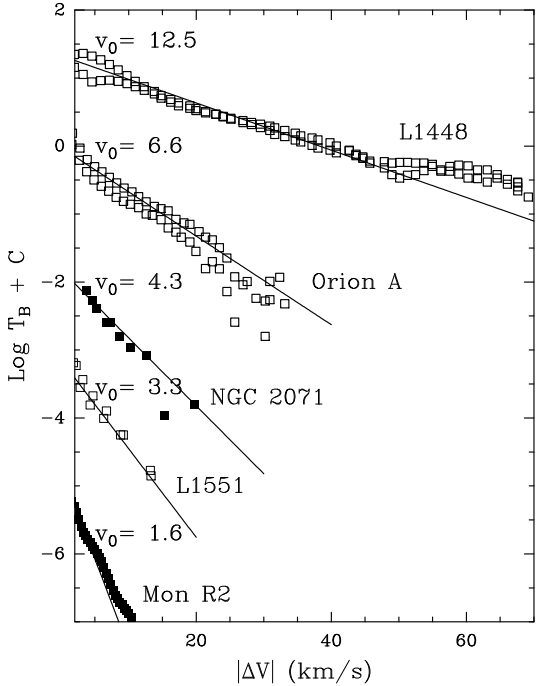}
\caption[]{Plots of the observed $I_{\rm CO}$($v$) in the $J$=2--1 line for L1448, Orion A, NGC2071, L1551, and Mon~R2 (open boxes and filled boxes; from Bachiller \& Tafalla 1999). For each source, the fit $I_{\rm CO}$($v$) $\propto \exp(-|v/v_0|)$ is drawn as a black line, and the value of $v_0$ is given in $\kms$. }
\end{figure}

Previous work described  the relation $I_{\rm CO}$($v$) in outflows by a broken power law,
$I_{\rm CO}$($v$)$\propto v^{-\gamma}$ with  $\gamma \simeq 1.8$ up to line-of-sight velocities
$v_{break} \approx 10-30 \kms$ and a steeper slope $\gamma \simeq$ 3--7 at higher velocities.
This behavior is successfully reproduced by jet-driven flows, as a result of CO dissociation above shock speeds of $20\kms$ and of the temperature dependence of the line emissivity (see DC03).  However, the underlying bowshock model predicts a power-law at low velocities on long time scales, rather than an exponential law, as observed in L1157-B1. It also predict a continuous range of temperatures in the swept-up gas, at odds with our finding that the L1157-B1 line profiles seem to be composed of three quasi-isothermal spectral components.

We show in Fig.~4 the CO $J$=2--1 observations of five outflows previously studied by Bachiller \& Tafalla (1999) and modelled by DC03~:
L1448, Orion A, NGC2071, L1551, Mon R2. We display in dashed the best fit to the data with a single exponential $I_{\rm CO}$($v$) $\propto \exp(-|v/v_0|)$. A very good agreement is observed in all cases (Fig.~4), with values of $v_0$ well in the range of those determined
in L1157-B1. We conclude that an exponential relation $I_{\rm CO}$($v$) $\propto \exp(-|v/v_0|)$ is a good approximation to the observed intensity-relation not only in L1157-B1 but in molecular outflows in general, with a reduced number of free parameters compared to a broken power law.

Our second main finding, that the exponential components in L1157-B1 appear quasi-isothermal and close to LTE, also has important implications. First, it shows that HIFI data are crucial to resolve ambiguities between sub-LTE vs LTE fits to CO excitation diagrams based on PACS data (Benedettini et al. 2012, Neufeld 2012). Second, it shows
that the CO flux up to $J_{\rm up}= 16$ is dominated in each component by the densest, coolest postshock gas.  Since such dense gas already reached a final constant speed, the broad velocity range of each exponential may require a broad range of view angles and/or shock speeds within the telescope beam. The exact origin of this spectral shape remains to be explained and may hold an important clue to entrainment and shock dynamics in molecular outflows.

\begin{acknowledgements}
We thank R. Bachiller and T.~Downes for providing us with the observational data presented in Fig.~3.
  HIFI has been designed and built by a consortium of institutes and
  university departments from across Europe, Canada and the United
  States under the leadership of SRON Netherlands Institute for Space
  Research, Groningen, The Netherlands and with major contributions
  from Germany, France and the US. Consortium members are: Canada:
  CSA, U.Waterloo; France: CESR, LAB, LERMA, IRAM; Germany: KOSMA,
  MPIfR, MPS; Ireland, NUI Maynooth; Italy: ASI, IFSI-INAF,
  Osservatorio Astrofisico di Arcetri-INAF; Netherlands: SRON, TUD;
  Poland: CAMK, CBK; Spain: Observatorio Astron\'omico Nacional (IGN),
  Centro de Astrobiolog\'{\i}a (CSIC-INTA). Sweden: Chalmers
  University of Technology - MC2, RSS \& GARD; Onsala Space
  Observatory; Swedish National Space Board, Stockholm University -
  Stockholm Observatory; Switzerland: ETH Zurich, FHNW; USA: Caltech,
  JPL, NHSC. C. Codella and C. Ceccarelli acknowledge the
  financial support from the COST Action CM0805 ``The Chemical Cosmos''. S. Cabrit
  and C. Ceccarelli  acknowledge the financial support from the french spatial agency
  CNES. G.Busquet is supported by an
  Italian Space Agency (ASI) fellowship under contract number I/005/007.
  B. Lefloch thanks the Spanish MEC for funding support through
  grant SAB2009-0011. J. Cernicharo thanks the Spanish MICINN for funding support
  through grants AYA2009-07304, and CSD2009-00038. Support for this work was provided by NASA  through an award issued by JPL/Caltech. The CSO is supported by the National Science Foundation under the contract AST-08388361.
\end{acknowledgements}

\begin{deluxetable}{crrcrrccccl}
\tabletypesize{\scriptsize}
\tablecaption{CO line observations. Telescope main-beam efficiency $\eta_{mb}$ half-power beam width, rms (per $1\kms$ velocity interval) and  integrated fluxes of the shock components $g_1$, $g_2$ and $g_3$ are given. Temperatures are expressed in $T_A{}^{*}$ units.}
\tablewidth{0pt}
\tablehead{
\colhead{Transition} & \colhead{Frequency} & \colhead{$E_{\rm up}$} &
\colhead{ObsID} & \colhead{$\eta_{mb}$} & \colhead{HPBW} & \colhead{rms} & \colhead{$g_1$} &
\colhead{$g_2$} & \colhead{$g_3$} & \colhead{Comment}\\
\colhead{} & \colhead{(GHz)} & \colhead{(K)} & \colhead{} & \colhead{} & \colhead{($\arcsec$)} & \colhead{(mK)} & \colhead{($\K\kms$)} & \colhead{($\K\kms$)} & \colhead{($\K\kms$)} & \colhead{} \\
}
\startdata
1--0 & 115.27120 & 5.5  & -          & 0.78 & 21.4 & 2.0 & 1.82 & 13.8 & 27.8  & IRAM \\
2--1 & 230.53800 & 16.6 & -          & 0.59 & 10.7 & 2.7 & 2.39 & 29.7 & 52.1 & IRAM  \\
3--2 & 345.79599 & 33.2 & -          & 0.65 & 22.0 & 130 & -    & 42.9 & 17.5 & CSO \\
5--4 & 576.26793 & 83.0 & 1342181160 & 0.75 & 37.4 & 8.0 & 2.58 & 39.8 & 8.4  & HIFI \\
6--5 & 691.47308 & 116.2& 1342207606 & 0.75 & 30.7  & 5.0 & 3.03  & 45.3 & 7.5&  HIFI     \\
7--6 & 806.65180 & 154.9& 1342201707 & 0.75 & 26.3  & 7.1 & 3.03  & 36.9 & 3.0&  HIFI   \\
     &           &      & 1342207624 & 0.75 & 26.3  &     &       &      &    &  HIFI\\
8--7 & 921.79970 & 199.1& 1342201554 & 0.74 & 23.0  & 10.0& 4.55  & 27.4 & -  &  HIFI\\
     &           &      & 1342207323 & 0.74 & 23.0  &     &       &      &    &  HIFI\\
9--8 & 1036.91239& 248.9& 1342200962 & 0.74 & 20.5  & 7.7 & 4.09  & 23.8 & -  & HIFI \\
     &           &      & 1342207641 & 0.74 & 20.5  &     &       &      &    & HIFI \\
10--9&1151.98544 & 304.2& 1342207691 & 0.64 & 18.4  & 36  & 3.79  & 9.61 & -  & HIFI \\
     &           &      & 1342196511 & 0.64 & 18.4  &     &       &      &    &  HIFI\\
13-12&1496.92291 & 503.2& 1342214390 & 0.72 & 14.1  & 46  & 4.55  & -    & -  &  HIFI\\
16-15&1841.34551 & 751.8& 1342196586 & 0.70 & 11.5  & 26  & 2.27  &  -   &  - & HIFI \\
\enddata
\end{deluxetable}

\end{document}